\documentclass{moriond}

\usepackage{hyperref}

\def\Journal#1#2#3#4{{#1} {\bf #2}, #3 (#4)}

\def\PLB{{\em Phys. Lett.}  B}
\def\PRL{\em Phys. Rev. Lett.}
\def\PRD{{\em Phys. Rev.} D}

\def\be{\begin{equation}}
\def\ee{\end{equation}}
\def\bea{\begin{eqnarray}}
\def\eea{\end{eqnarray}}

\newcommand{\Photo}{}

\begin{document}

  \begin {flushright}
    TTP22-035, P3H-22-057 
  \end{flushright}

\vspace*{1cm}
\title{Theoretical summary of Moriond 2022: QCD and high-energy interactions}

\author{Kirill Melnikov }

\address{ Institute for Theoretical Particle Physics, Department of Physics,
  Karlsruhe Institute of Technology, \\
76128, Karlsruhe,   Germany
}

\maketitle

\abstracts{
  I review   theoretical talks presented at the session on QCD and high-energy interactions of the Moriond 2022 conference.
}

\section{Introduction}

The 56th edition of Rencontres de Moriond  held earlier
this year  was one of the first in-person conferences after almost two years of the pandemic  isolation during 
which many interactions were limited to an on-line format. In fact, this was the very
first in-person conference for me since a very long
time.   Although we all learned to appreciate the convenience
of on-line conferences and meetings,  participants' eagerness to talk to each other  \emph{in person} about
physics was clearly seen  at Moriond. 

Theory talks at  ``QCD and high-energy interactions''
session of Moriond covered many areas of contemporary particle physics often going beyond
a conventional understanding of what ``QCD Moriond'' is all about.   This
is a great feature of this conference as it emphasizes the
unity of particle physics,  including  its research goals and methodologies.

Particle physics  is defined by big questions that
it tries to answer. These questions are well-known; they
include unification of known interactions, an underlying cause of electroweak symmetry breaking,
origin of families, fermion masses and Yukawa couplings in the Standard Model,
an asymmetry between visible matter and anti-matter,
nature of dark matter, role of gravity and its  connection to  other  known interactions etc.

It is well known that  progress in addressing these
questions has been quite  slow and, clearly,  not for the lack of trying.   However, continuous 
attempts  to study them  challenge   experimental
and theoretical status quo  in  particle physics,  and   push  the scientific frontier  into unchartered
territories.  As the result,  cases are often  encountered 
where things do not work as expected and where tensions between theoretical expectations and experimental results
become obvious.  Although we often refer to such  cases as the ``anomalies'',
they are perhaps better viewed as the   ``growing  pains'' needed to get to a new level of understanding
of fundamental laws of Nature.  And, in spite of the fact that, time and again,  a 
better mastery  of the Standard Model, rather 
than the discovery of  physics beyond it, emerges as the result
of painstaking investigation of the anomalies,   steady
progress that is driven by points of content in particle physics  is not to be overlooked. 

  Current anomalies in  particle physics naturally became the focus points of the conference, with 
  flavor anomalies, lepton non-universality and   the muon anomalous magnetic moment dominating the discussion. In addition, we 
  heard about   new theoretical developments in QCD,  an  interplay between precision physics at the LHC and searches for
   physics beyond the Standard Model, as well as   complex aspects of QCD dynamics. 
   I will briefly review all the different theoretical talks delivered  at the QCD Moriond 2022
   starting with the discussion of the muon magnetic anomaly.

\section{Muon magnetic anomaly}

The puzzle  of the muon anomalous  magnetic moment
has been with us for more than twenty years by now \cite{ben1} and recently significant,
${\cal O}(4 \sigma)$, discrepancy between theoretical predictions \cite{Aoyama:2020ynm}
and experimental measurements \cite{ben1}
has been  confirmed by the first result \cite{fnal1}  of the FNAL experiment.  
The difference between measured and expected
 values $a_\mu^{\rm th} - a_\mu^{\rm exp} = 251(59) \times 10^{-11}$ is quite large.  In fact, it is
a factor of two larger than the four-loop QED contribution to $a_\mu$,  
more than fifty percent larger than the one-loop electroweak
contribution  to $a_\mu$ and  is about twice as large as the so-called hadronic
light-by-light scattering contribution to $a_\mu$.
On the contrary, it is just about   $3.5$ percent of the hadronic vacuum polarization contribution to $a_\mu$,
making good understanding of $a_\mu^{\rm hvp}$ a very important issue.

Given the many existing checks of the  QED and electroweak contributions as well as obvious difficulties with the
theoretical
description of low-energy hadron physics, current work on resolving the discrepancy focuses on
the hadronic vacuum polarization
and on the hadroninc light-by-light scattering contributions to the
muon magnetic anomaly.  Although these two hadronic contributions are mentioned in a single
sentence, they are actually quite different. Indeed, as noted  above,
the hadronic vacuum polarization contribution is large and needs to be known
to a few-percent precision whereas the hadronic light-by-light scattering contribution
is rather small and knowing it to about twenty percent precision
is sufficient. This difference between the two hadronic contributions is reflected  in the way 
they are currently dealt with and discussed. 

Let us  start with the  hadronic vacuum polarization. For the past sixty
years, the standard way to compute it  was to use the dispersion representation for 
the vacuum polarization function  relating it to the measured cross section of $e^+e^-$ annihilation to hadrons.
Using  experimental data for  $\sigma(e^+e^- \to {\rm hadrons})$, the dispersion integral is calculated  numerically leading
to a precise result for the hadronic vacuum polarization contribution to the muon magnetic anomaly.  This standard approach
was discussed in great detail by Zhang \cite{zhang}.

Although there are various caveats and difficulties related to the computation of the dispersion
integral, including reliability of data, consistency between
various experiments and treatment of radiative corrections,   understanding its oder-of-magnitude is
quite straightforward \cite{book}.  Indeed, all one needs to do  is to account for  three lightest spin-one hadronic resonances  
$\rho$, $\omega$ and $\varphi$ in the dispersion integral treating them as narrow peaks with
known  masses and branching ratios to leptons. Supplementing their contribution to the dispersion integral
with a  continuum contribution to accommodate 
physics beyond the center-of-mass energy of ${\cal O}(1)~{\rm GeV}$,  one obtains
the result \cite{book} that accounts for about \emph{ninety  percent} of $a_\mu^{\rm hvp}$. 

Of course, the problem with the
hadronic vacuum polarization contribution  is that  it  is so large ($a_\mu^{\rm hvp} \approx 7000 \times 10^{-11}$),
that knowing it to ninety percent is insufficient. To match the current experimental uncertainty, 
we need to understand it to better than a percent and reaching this precision is very challenging.  Figuring out how to do this in a controllable
way  was the focus point of the
practitioners of the dispersion-relations method in the
past; these efforts were reviewed by Zhang \cite{zhang}.  In fact, in spite of a few open questions and tensions
(e.g. KLOE data vs. BABAR data), it does appear that \emph{no} foreseeable modification of either data or analysis methodology 
can shift $a_\mu^{\rm hvp}$  beyond the estimated uncertainty  ($ \sim 40 \times 10^{-11}$)  \cite{zhang}. 

However, although the most precise results for the hadronic vacuum polarization
were traditionally obtained with the help of dispersion relations
and experimental data, the situation has changed in the past few years.
Indeed,  in 2020 the BMW collaboration published a lattice computation
of the hadronic vacuum polarization contribution to
the muon magnetic anomaly which is claimed to have a sub-percent precision. This calculation
was presented by Szabo \cite{szabo}.  In a remarkable twist of fortunes for SM physics, the new
calculation is  about $140 \times 10^{-11}$ larger,
than estimates of the hadronic vacuum polarization obtained with the  dispersive  method and,
although this shift represents just about $2 \%$ of the total hadronic vacuum
polarization contribution to the muon magnetic anomaly,  it wipes out more than
50 percent (!)  of the discrepancy between theory and experiment reducing  it to a meager   ${\cal O}(2\sigma)$.
Hence, if correct, this result  will be responsible for the major change in the muon magnetic anomaly  story,
forcing  a prospective  ``harbinger of New Physics''
to become   ``a poster child of the Standard Model''.

However, we are not there yet since at this point 
there is no reason   to trust the new
lattice computation more than the results of the dispersive approach. The challenge, therefore, is to understand where and why
the discrepancy between these two very different computations comes from.    This question can be addressed in several ways.
    First, the result of the BMW collaboration should be confirmed or refuted by other lattice collaborations
  and this applies to both the central value and the uncertainty estimate. 
    Second, lattice and dispersive computations need to be compared. This is not easy because 
   lattice operates   in the Euclidean space and, therefore,  $\sigma(e^+e^- \to {\rm hadrons})$ cannot be computed on the lattice.
  However, it is clearly possible to  compute various  \emph{moments}
  of the hadronic vacuum polarization function  using both the
  dispersion method and lattice QCD,  and to construct these moments 
  in such a way that  contributions of particular  energy  intervals  are strongly emphasized.
  If such a comparison is performed systematically, it  should allow us to eventually 
  pinpoint the   energy intervals that are resposible for the discrepancy. This will be the most welcome development
  because   such
  an understanding will have important implications beyond the muon magnetic anomaly. For example, 
  both lattice and dispersive  methods can be
  used to compute the fine structure constant  $\alpha(M_z)$  whose 
  value is highly relevant  for the  precision electroweak
  fit. 

  Another contribution to the muon magnetic anomaly
  which for a long time was considered to be the main ``troublemaker''  is the hadronic light-by-light scattering
  contribution. One of the main reasons it got this  status was a sudden change in the  \emph{sign} of
  this contribution that led to its dramatic, 
  overnight increase \cite{raf} from ${\cal O}(-100) \times 10^{-11}$ to ${\cal O}(+100) \times 10^{-11}$.
  Although the  sign issue was later clarified
  to be an  isolated  incident which, at its core, proves that the FORM manual \cite{form} is not an exciting read,  the mistrust towards this contribution remained.

  However, it has to be recognized that  \emph{already} since 2002 
  the hadronic light-by-light scattering contribution has been  estimated to be close to $100 \times 10^{-11}$, with an uncertainty of about $20 \times 10^{-11}$, which
  is about a half of the estimated uncertainty in $a_\nu^{\rm hvp}$ as discussed above. 
   The point of view  that the
  hadronic light-by-light scattering contribution is actually
  fairly well understood  was presented in the talk by J.~Green \cite{green} who discussed the new lattice
  computation of $a_\mu^{\rm hlbl}$. 
  Their  result,  $a_\mu^{\rm hlbl} = 106(16) \times 10^{-11}$, is in agreement with many earlier estimates of this quantity
  obtained using a variety of phenomenological method.  And, as a further illustration of the remarkable consistency
  of theoretical predictions for $a_\mu^{\rm hlbl}$, 
  I cannot help but mention that its central value is identical to the 
  result \cite{glsgo} of the so-called ``Glasgow consensus'' by J.~Prades, E.~de~Rafael  and
  A.~Vainshtein, published already in  2009. 

  I would like to emphasize that the new lattice result reported in Green's talk \cite{green}
  is very important as it   provides further support to an understanding, that has been  emerging for several years by now,  
  that presumptive lack of theoretical control  of the
  the hadronic light-by-light scattering contribution
  to the muon $g-2$   cannot be the \emph{only}  reason for the discrepancy of the
  muon magnetic anomaly. Whether  the $a_\mu$ puzzle  will eventually be 
  resolved  by  a  big problem in $a_\mu^{\rm hvp}$, or by a collection of small(ish) deviations in $a_\mu^{\rm hvp}$, $a_\mu^{\rm hlbl}$ and
  in the experimental value of $a_\mu$ which  all work just in the right way,  or by a New Physics contribution,  is impossible
  to say now and remains to be clarified  in the future. 

  \section{Flavor physics}
  
Given the fact that some of the most persistent and unusual discrepancies between theoretical expectations and experimental results
currently occur in flavor physics,  talks  on flavor physics were an  exciting  part of the conference. 
The discussion of flavor physics  started with an overview talk on the current status of the anomalies 
by M.~Neubert \cite{neubert}.  Clearly, as with any anomaly,   a  proper evaluation of the situation requires
good  understanding of the quality of the Standard Model  predictions, ideas about a possible explanation of the observed
discrepancies by New Physics and a good overview of how  such ideas  fit into a global picture of checks of the Standard Model.

Two talks at the conference addressed the quality of   the SM predictions in $B$-physics.
Capdevila reported \cite{capdevilla}   a refined extraction of the CKM matrix element $V_{cb}$ from inclusive semileptonic
$B$-decays into final states with charm quarks.   The refinement described by Capdevila  
comes from the inclusion of the  recent   N$3$LO QCD  calculation of the partonic rate for $b \to c$ transition 
\cite{stein} into a theoretical prediction for $B \to X_c l \nu_l$
decay rate computed within the heavy quark expansion. 
If  the N$3$LO QCD corrections are  included into the analysis,
the central value of $V_{cb}$ remains practically unchanged when compared to an earlier NNLO QCD analysis, 
 but the  uncertainty in the extracted value is reduced by about twenty five percent. 

 This  is  both  good and  bad news. It is a good news because it shows  that the extraction of $V_{cb}$ from inclusive
 $B$-decays is theoretically robust. However, it is also a bad news because
 it leaves the discrepancy between exclusive and inclusive $V_{cb}$ 
 measurements unchanged, at the level of one to three standard deviations. 

 Gubernari \cite{gubernari} discussed the Standard Model  predictions for the exclusive $B$ decays into a strange
 meson ($K,\varphi$) and a lepton pair.
 Making  accurate  predictions for \emph{exclusive}  decays  is  difficult in general; in case of $B \to Kl^+l^-$ and
 $B \to \varphi l^+l^-$ decays ,  the problem
 is enhanced because of the so-called charm loop contributions that are  notoriously
 difficult to treat reliably.  By using a clever combination of the operator product expansion
 and dispersion relations, Gubernari showed how to derive upper bounds on these complicated  amplitudes. 
  However,  even after this theoretical refinement,  the significant tension between the SM predictions
  and experimental measurements  remains. In particular, is does not appear that
  a large discrepancy between measured and expected values observed in the distribution of the so-called $P_5$ observable 
   can be attributed to the deficiencies  of  the SM predictions. 

   Given that the Standard Model predictions seem to be   robust, it is important  to
   discuss flavor  anomalies from  the perspective of beyond the Standard Model (BSM) physics. 
    There were four  theoretical talks on this subject and  the unified   message of these talks 
 was that  these anomalies can certainly be early manifestations   of New Physics  which, so far, could have avoided
  detection at the LHC in spite of the incredible amount of data collected there  during the first two runs. 
  This is a reassuring  message that  emphasizes  the complementarity
  of different ways to probe Nature and the  importance of  low-energy  physics
  as a path finder for the LHC.

  The four talks on the BSM explanation of flavor anomalies showed that there are many different ways to accommodate
  them in a consistent framework.
  Allanach presented a particular  $Z'$ model \cite{allanach}    that explains lepton non-universality, 
  does not (significantly) spoil precision electroweak fit and
  is practically unconstrained by the current searches for $Z'$'s at the LHC. 
    Iguro \cite{iguro} pointed out that, due to relaxed   constraints from the decay $B_c \to \tau \nu$, the charged-Higgs
    explanation of the $R_{D^*}$ anomaly becomes possible and, if the charged Higgs boson is relatively light $140~{\rm GeV} < m_{H^-}
    < 400~{\rm GeV}$,
    there is no contradiction with the current LHC bounds.  He then noted  that this
    explanation of the $R_{D^*}$ anomaly can be tested at the LHC already with the existing data 
    provided that one uses the  production of the Higgs boson in  association
    with a $b$-jet, $g \bar c \to H^-(\tau \nu_\tau) b$,  to suppress the 
    large background from the Drell-Yan process $pp \to \tau \nu_\tau$.

 Boussejra \cite{bousseja} pointed out that SUSY models with non-minimal flavor violation can also explain flavor
  anomalies without getting into a conflict with other precision observables and the negative results from LHC searches. 
  Finally,   Crivellin \cite{criv} emphasized the importance of taking a  high-level view on flavor and related
  anomalies since many of them can actually be linked to each other  within  BSM models. 
  To illustrate this point he presented a model  that simultaneously explains the so-called Cabibbo anomaly,
  the old $Z \to b \bar b $ anomaly and the $\tau \to \mu \nu \nu$ anomaly which, at first sight, do not need
  to be related.

  It is well-known that $B$-physics  observables are computed starting from an effective Hamiltonian which
  involves effective operators weighted with their Wilson coefficients. These Wilson coefficients have particular values
  in the Standard Model, so that many of the flavor anomalies can be thought of as
  differences in their measured 
  and  expected SM values.   To explain an  anomaly
  with a BSM theory, it is essential to    re-compute the Wilson coefficients of operators in the effective
  Hamiltonian in a new theory.
  Although it is very well understood how to do this, 
   the bookkeeping is challenging.  Santiago presented 
  a tool, dubbed Matchmaker, which allows one to automatically compute the Wilson coefficients by matching  new BSM 
  theory to an effective field theory through one-loop \cite{santiago}. 
  One can hope  that the availability of such a tool will be helpful for  further exploration of
  the theory space using 
  experimental data on $B$-decays.

  \section{Dark matter} 

  Dark matter was not the major focus of the QCD Moriond,  but there were several talks on this exciting topic. 
  White described  \cite{white} a comprehensive global fit for a Dirac fermion dark matter,  that also
  showcased  impressive capabilities of  the GAMBIT program \cite{gambit}.  White's message was that 
  although significant parts of the parameter space can be excluded, the Dirac DM is still
  a  viable option.

 Rolbiecki pointed out \cite{robliecki} that it is possible to use the monojet  signature  to improve constraints
  on electroweakino dark matter in a situation where mass splittings between electroweakinos are small. 
  A standard experimental analysis to explore  such a scenario employs the monojet signature. However, as  Rolbiecki
  explained,  stronger constraints are obtained by recasting  searches for squarks and gluinos, where   ``jets + missing energy''
  signatures  are also studied, into bounds on the electroweakino DM. 

  Finally, Ruderman \cite{rudermann}   described a new mechanism  of how the dark matter density can be generated in the early
  Universe.   He pointed out that,  in case there is an interaction term in the Lagrangian
  between three dark matter particles and one   Standard Model particle,  a new collision term  appears in the Boltzmann
  equations that forces   DM density to grow exponentially.
  This new mechanism   changes the famous relation between the DM  equilibrium density and the annihilation 
  cross section between dark matter particles, so it is useful to be aware of the fact that
  there are cases where such a relation does not hold and a very different mechanism for producing the DM density is at play. 
  
  \section{Collider physics}
  
  Development of theoretical methods that can be used to describe hard scattering processes at colliders
  has accelerated in recent
  years \cite{gudrun}. In particular, the appearance of robust  subtraction and slicing schemes for
  higher-order perturbative computations, as well
  as advances in technologies for computing multi-loop amplitudes, 
  resulted in a number of impressive results for basic collider processes obtained  recently. 
  Some of these results were presented  at QCD Moriond. 

  Perhaps the best-known process that occurs at a hadron collider is the 
  Drell-Yan process $pp \to l_1 l_2 +X$.  Depending on the final
  state, it  is facilitated by a neutral or by a charged current.  Studies
  of the Drell-Yan processes are very  important for the LHC phenomenology; they include  such physics topics
  as  properties of $Z$ and $W$ bosons,  parton distribution functions, 
  calibration of the detectors, searches for BSM physics and more. 

  It is therefore not surprising that   theoretical studies of these processes are extremely advanced.
  Just \emph{how} advanced they are became evident from the  three talks on  the cutting-edge computations
  of higher-order  perturbative  corrections to the Drell-Yan process.   Yang \cite{yang} described the calculation   
  of N$3$LO QCD corrections to the rapidity distribution of a vector boson,  while Rottoli \cite{rotolli}
  presented a computation of fiducial cross sections of the Drell-Yan processes at the same perturbative order. 
  Finally,
  Buonocore \cite{buonocore} discussed  the  calculation  of mixed QCD-electroweak corrections to the neutral-current-mediated
  dilepton production  both
  at the resonance and in  the high-invariant mass region of a dilepton pair.
    The bottom line of these tour de force   computations seems to be that  various observables
  in  the Drell-Yan process can be described with an astounding  precision  of about 1-2 percent.

  It is worth pointing out  that N$3$LO QCD corrections to observables
  in the DY processes    turned out to be not much smaller than the 
  NNLO QCD ones \cite{yang,rotolli} and, therefore,  larger than expected. This fact
  was already noted when calculations
  of N3LO QCD corrections to the total cross sections of $Z$ and $W$ production at the LHC
  appeared \cite{mist1,mist2}.  The explanation of  \emph{why} this happens  is
  probably multi-facet.   At least  partially, it may  be attributed to the fact that NNLO QCD 
  corrections to the Drell-Yan processes are probably smaller than they should be because of 
  accidental cancellations between contributions of different partonic channels. Another reason can be that 
  parton distribution functions at  N3LO QCD are not yet known.

  Clearly, it is important to go beyond admiring the technical wizardry behind  these computations and to figure out  how to use 
  higher precision of theoretical predictions 
  to learn more about physics.  Although  full appreciation  of the opportunities that the new
  results \cite{yang,rotolli,buonocore} open up  is 
   still to come, first glimpses of what can be expected could already be seen at the conference.  For example, 
   Schott \cite{schott} pointed   out that a  very competitive value of the  strong coupling constant,
   $\alpha_s(m_Z) = 0.1185 \pm 0.0015$, can be obtained 
   from the  transverse momentum spectrum  of the  $Z$ bosons. In fact, Schott's analysis does not
   include all available perturbative
   corrections to the $Z$-boson  transverse momentum distribution so that, in principle,
   it can be be further refined if necessary.

   Another interesting opportunity, described by Scimemi \cite{scimemi}, is to use data on the Drell-Yan process, together
   with high-precision theoretical predictions, 
   to extract transverse-momentum dependent (TMD) parton distribution functions.
   These functions are needed  to descirbe the $p_\perp$-distribution
  of a vector boson   at very low $p_\perp$  where tiny  transverse momenta of colliding quarks
  cannot be neglected.     Scimemi pointed out that because of  the factorization theorems,  there is
  an intimate relation between TMD PDFs and
  regular PDFs  which implies that any uncertainty intrinsic to collinear PDFs gets transferred to an uncertainty in  TMD PDFs.
  He also argued  that it is very important to account for flavor dependences of TMD  PDFs in the global  fit and that
  if one does that, the consistency of TMD PDFs extracted from various data sets significantly increases.

  It is intuitively clear that the   availability of more precise  SM results should allow for  stronger constrains on  BSM contributions
  to basic hard processes such as  Drell-Yan,  and  Giuli \cite{giuli} described an  example of this in his talk. 
  He considered the  case of a  heavy and relatively broad resonance contributing to the dilepton spectrum  
  and pointed out that  it is  difficult  to detect it using  conventional bump-hunting methods. 
  He then pointed out that one can observe  shape modifications in the dilepton invariant mass distribution caused by a broad resonance
  provided that there is a 
  good control of parton distribution functions at high values of the Bjorken $x$,  and that  high-precision
  QCD predictions for the dilepton invariant mass spectrum are available. 
    He argued that by  including data on  the charge asymmetry and the forward-backward asymmetry  into a simultaneous fit for 
  high-$x$  parton distribution functions and a prospective resonance contribution to the dilepton invariant mass spectrum,
  reach for broader and heavier resonances improves. 

  LHC physics requires good understanding of final states with QCD jets and there
  were a few talks at QCD Moriond that described recent advances in this endeavor. 
  Poncelet reported \cite{poncelet} on a computation of NNLO QCD corrections to 3-jet production at the LHC.
  This is a very impressive   result that further highlights an enormous technical  progress that occurred in the
  field of perturbative QCD in recent years. 
  Without a doubt,   this tour de force calculation,
  one of the most complicated calculations in perturbative QCD ever performed,  will 
     find many phenomenological applications in the future, including the determination of the strong coupling constant and refined 
     studies  of jet dynamics.
     
     Definition of jets requires certain
     prescriptions to combine energies of  various particles into directional energy flows; these prescriptions are known
     as ``jet algorithms''.  Over time 
     jet algorithms evolved  \cite{salam} from experimentally-convenient but
     theoretically-problematic seed cone algorithms to modern ones, such as e.g.
     the anti-$k_\perp$ algorithm, which is   practically  impeccable from both
     the experimental and theoretical points of view.
     However, even if a great  solution is available, one can always try to do  better. 
     In this spirit,   Cerro discussed  \cite{cerro} a new algorithm for clustering hadrons into
     jets based on machine learning methods. A  few
     examples were shown which demonstrated that in certain cases
     the new algorithm outperforms the conventional ones. 
     It is clear that many more studies are needed before this algorithm will get close to becoming  as widely accepted
     as  the   anti-$k_\perp$ one, but it will be  interesting to watch how this story develops further. 
  
Dreyer \cite{dreyer} discussed resummation of the so-called non-global logarithms for QCD observables at lepton colliders. 
Non-global logarithms represent a particular class of  enhanced contributions to observables
that   appear  if radiation to certain phase-space regions is restricted.
The problem of the resummation of non-global logarithms
is known to be rather difficult
but significant simplifications occur if it  is studied in the large-$N_c$ limit.
In fact, in this case,  resummation of non-global logarithmic  contributions at next-to-leading-logarithmic accuracy 
can be performed  using a relatively simple extension of the so-called Banfi-Marchesini-Smye equation \cite{bms}.
Progress with  the understanding of non-global logarithms and their resummation reported by Dreyer is a welcome development since
it  is  essential for designing  parton showers with the 
next-to-next-to-leading logarithmic accuracy.

We now change gears and talk about physics of top quarks which  was also discussed in theoretical talks at Moriond QCD. Top quark
physics is a big  part of  the LHC research program since the  LHC is a
top quark factory. This  fact enables  detailed studies of  various processes with top quarks, including
searching for possible contributions of physics beyond the Standard Model,
and the exploration of  top quark  properties.

  Jezo \cite{jezo} pointed out that  \emph{off-shell} contributions to signatures that are used to identify top quark pair production
  and study top quark properties may be important.
  Although the importance of these contributions  -- or lack of it -- must depend on details
  of experimental analyses,  there is a theoretical aspect that is worth emphasizing. Indeed, 
  the development of computational  methods during the past decade
  resulted in a situation where, for many processes,
  it is more straightforward (if not simpler) to provide
  predictions for $pp \to bWbW+X$ final states than to split them into
  double-resonant, single-resonant and non-resonant
  contributions. This fact further implies that the relative importance of various contributions, including signal-background
  interference etc., are decided by experimental constraints imposed on a \emph{single} theoretical calculation,
  and do not require complex, poorly justified constructions
  that were used earlier to combine the different contributions. 
  
  However, it is not always possible to pursue this program since for certain final states complete computations
  remain    too  complicated even with  modern methods.  According to Jezo, this happens when 
  $W$-bosons in top decays are allowed to decay \emph{hadronically}.  To make progress, Jezo proposed  to consider \emph{on-shell}
  $W$-bosons since in the $\Gamma_W \to 0$ limit decay products of $W$-bosons cannot interact with other
  parts of the process by  QCD exchanges. Because of this,
  a connection  between corrections to  fully-leptonic and semileptonic signatures in
  the off-shell $t \bar t$ production process arises and the computation of QCD corrections in the semileptonic case
  simplifies enormously.

  Devoto discussed a calculation of the $t \bar t$ pair production at the LHC  with the NNLO QCD accuracy \cite{devoto}.
  So far this calculation is performed for stable top quarks and in this limit 
 it confirms earlier pioneering computations of Czakon  et al. \cite{czak}.
 This is  a strong check on both calculations because 
 Devoto's calculation  employs a
 different method  to combine separately-divergent real emission contributions and virtual corrections. 
   Another interesting point mentioned by  Devoto   is that
this computation will become part of a new release of a publicly-available program  MATRIX \cite{matrix}  which
is quite useful since so far there is no public program for calculating  kinematic
distributions for top quark
  pair production in hadron collisions with the NNLO QCD accuracy.

  I believe that when a typical  participant  of QCD Moriond
  thinks about measurements of the top quark and the Higgs boson masses,
   collider, rather than cosmological, aspects
  of this problem come to mind.    However, it is well known that the precise knowledge
  of these masses  has very  important implications for the stability of electroweak vacuum \cite{stab}.
  Santos  \cite{santos} reminded us about this connection in his talk, albeit in a slightly different context.
 He pointed out that under certain circumstances uncertainties in $m_t$ and $m_H$ may  preclude 
 an  interpretation of  the signal of gravitational waves   that originate in the course of
 the first-oder phase transition  in  the early Universe. Santos' observation provides yet another  motivation for 
 measuring  the top quark mass more precisely and
 advanced theoretical  predictions for top quark production cross sections and kinematic distributions play a particular
  important role in this endeavor.

  We have seen yet another example of the creative  use of high-precision theory in a talk by Altakach \cite{alkatach}
  where he discussed
  a possibility to constrain BSM physics by comparing fiducial cross sections with high-precision predictions
  for signals and backgrounds.   He considered   a particular model of New Physics
  where a $Z'$ boson  couples to quarks of the first and the third generations but not to leptons.
  This  $Z'$ is also relatively broad so that it is not possible to discover it by searching   for  a peak  in the $t \bar t$
  invariant mass distribution.  Altakach  showed  that   
  by improving theoretical  predictions for the signal process and the various backgrounds, one indeed 
  obtains  better bounds on the $Z'$ mass. However, these bounds 
  are still somewhat worse \cite{alkatach} than  the bounds on the parameters
  of this model that are obtained from dedicated experimental searches,  presumably because
  kinematic information,  rather than just fiducial cross sections,  is used there.

Moving beyond  the top quark physics,
Neuwirth \cite{neuwirth}
reported on a study of NLO QCD corrections to squark-gaugino production at the LHC supplemented
with the soft gluon resummation. He showed
that the soft-gluon resummation,  consistently combined with
the NLO QCD computation, increases the cross section
  by about six percent and reduces the QCD scale uncertainty to about 5 percent.
  Hopefully these results will  be used to further constrain the parameter  space
  of supersymmetric  models or, perhaps, help to infer properties of squarks and gauginos if  they are finally discovered. 

\section{Parton distribution functions}
  
  Predictions for hadron collider physics are impossible without knowledge of parton distribution functions (PDFs).
  Learning about them requires complicated machinery that is usually managed 
  by highly-specialized collaborations.    Two theoretical talks on  the issue of PDFs' extractions
   which addressed  unorthodox aspects of PDF physics were presented at Moriond.

  Magni \cite{magni} argued that  LHCb data on $Z$+charm production show
  evidence of  non-perturbative component of the charm PDF in the proton.
  This, of course, is an interesting result; discussions about whether or not there is an ``intrinsic charm'' in the proton
  have been going on since quite a long time \cite{brod}  so that  a confirmation of this idea would be illuminating.  However, 
  at this point it does not appear that a decisive  conclusion about this matter is possible
  since the evidence is rather weak, about one sigma \cite{magni}.  Moreover,
  the non-perturbative component by itself is not  large;  according to the
  analysis by Magni   \cite{magni}  the intrinsic charm carries about
  one percent of the proton momentum and (not surprisingly) the intrinsic-charm PDF
  is peaked at the large $x$-values.
  Given all that, it can be expected that
  reaching  a definite verdict on the issue of the intrinsic charm in the proton
  will probably be difficult, 
  but it is quite interesting that there are attempts to make progress in that direction.  
  
  Bertone \cite{bert} discussed uncertainties in the extracted value  of the strong coupling constant
  and  parton distributions   due to the 
   imprecise knowledge of the anomalous dimensions and $\beta$-functions  as well as simplifications made 
   in the renormalization group equations that allow one  to solve them analytically.
   Bertone  finds that if PDFs are defined at low scales and then computed at an electroweak scale by solving
   Altarelli-Parisi equations, such truncations may lead to errors of a few  percent.
   To be sure, percent-level uncertainties in PDFs are not unheard of  but it is clearly  important to identify \emph{all} sources
  of such uncertainties including  pure theoretical ones. 
  
\section{Unorthodox topics} 

Often, it is not too difficult  for an experienced person to predict topics which  will be discussed at a conference dedicated
to QCD and high-energy interactions.  This is a reflection of the  fact that the scientific progress 
is continuous, at least  most of the time.  However, at every conference there are  a few
talks beyond the  expected  narrative and such
talks often become some of the most interesting ones simply because they are unexpected.
A number of such talks were presented at QCD Moriond this year.  

Zanderighi \cite{zanderighi} pointed out that one can find many things in the proton, including leptons and photons.
  In a way, this  observation should extend the notion of the complimentarity of lepton and proton colliders -- those of us who thought
  that LHC was colliding quarks and gluons will probably have  to think again!   
  Of course, the fact that one can find electrons or photons in a proton is not very surprising
  since, rhetoric aside, one  simply talks  about high-order
  QED processes. What is perhaps unexpected is that \emph{rates} of such  processes are strongly
  enhanced because of the kinematics of the quasi-collinear splittings. The  result of such an enhancement is that effects
  that could have been neglected completely become somewhat
  relevant and can often be best described by using the notion of distribution functions of leptons and
  photons in a proton.

  Zanderighi  explained \cite{zanderighi}  that one can determine photon and lepton PDFs from  the
  known structure functions measured in deep-inelastic scattering,   and use this information
  to set up theoretical description of QED-initiated processes  retaining full information
  of the events' kinematics, which is
  important for the LHC physics. 
  She also showed a few examples where  presence of leptons and photons in the proton becomes somewhat
  of a game changer.   For example, a
  significant lepton component of the proton leads to  additional contributions to leptoquark production
  in proton proton collisions,   and sometimes  these contributions are large
  enough to affect the exclusion limits in a significant  way \cite{zanderighi}.

Schott \cite{schott} proposed to search for instantons at the LHC.    Instantons are QCD  field
  configurations that describe a transition between two non-equivalent QCD vacua. It is expected that the instanton-production 
  process can manifest itself through a production of a spherically-symmetric multi-particle final state
  with certain polarization features of final-state partons.   Unfortunately, it is very
  difficult to predict the production cross section for instantons in hadron collisions and
  it is also hard to say if imprints of quark polarization can be observed in features of
  mesons and baryons that are actually measured.   At any rate,  this unorthodox proposal is quite interesting
  and it clearly  provides a new motivation for   many people \cite{workshop}  to think about an  old  problem, namely how to observe
  quasi-classical solutions,  present in non-abelian gauge theories,  experimentally.

  Tantary \cite{tantary}
  described a perturbative  calculation of a free energy in $N=4$ SUSY Yang-Mills theory.
  Interestingly, this computation can be used to reconstruct the exact free energy function since 
  in this theory the free energy  can also be computed    in the non-perturbative regime using  gauge-gravity duality.
  Extrapolating between perturbative and non-perturbative regimes
  can be done using Pade approximation. 
  Higher-order perturbative computations, performed  by Tantary, show convergence towards the
  Pade result constructed using the information at  lower orders.    Hence, we have an example of a theory where the free energy is known {\it exactly},
  for any value of the strong coupling. It is an interesting
  question what, if anything,
  can be learned from this result for thermodynamics of QCD plasma; at this point, I don't think
  there are clear ideas about this.

\section{The future}

It is peculiar   that the \emph{very first theoretical   talk} at  QCD Moriond 
provided an outlook on the future of collider physics \cite{rob}. Perhaps, this simply shows how urgent this matter is,
as multi-decade planning for future facilities is a commonplace in particle physics.  Given this, it is important to ask
how to  optimize the process of moving forward and  to obtain the richest  outcome in  terms of physics  in the shortest amount
of time.   Franceschini \cite{rob} provided many instructive examples and interesting considerations comparing the physics reach of different  colliders  that are being
discussed as potential  successors to the LHC. 

 \section*{Acknowledgments}
 I am grateful to the organizers of Moriond QCD and high-energy interactions for the invitation  to give this talk.
 This research is partially supported by the Deutsche
Forschungsgemeinschaft (DFG, German Research Foundation) under grant
396021762 - TRR 257.

\end{document}